\documentclass{elsart}
\usepackage{natbib}

 \usepackage{epsfig}

\usepackage{amssymb}

\def\la{{\lesssim}}
\def\ga{{\gtrsim}}

\newcommand{\bea}{\begin{eqnarray}}
\newcommand{\eea}{\end{eqnarray}}
\newcommand{\beq}{\begin{equation}}
\newcommand{\eeq}{\end{equation}}

\def\apj{{ApJ}}

\begin{document}

\begin{frontmatter}
\title{Whence particle acceleration}

\author[mm]{Mikhail V. Medvedev},
\author[as]{Anatoly Spitkovsky}
\address[mm]{Dept. of Physics and Atronomy, University of Kansas, 
Lawrence, KS 66045}
\address[as]{Dept. of Astrophysical Sciences, Princeton
University, Princeton, NJ 08544}

\begin{abstract}
We discuss how the electrons in relativistic GRB shocks can reach 
near-equipartition in energy with the protons. We emphasize the non-Fermi
origin of such acceleration. We argue that the dynamics of the
electrons in the foreshock region and at the shock front plays 
an important role. We also demonstrate that PIC simulations can
directly probe this physics in the regimes relevant to GRBs. 
\end{abstract}

\end{frontmatter}


{\bf Electron acceleration/heating} ---
There is a lore that charged particles are accelerated at shocks
by the Fermi mechanism. Numerical simulations show that 
although Fermi accceleration may be present, it cannot heat
the bulk electrons to near-equipartition with the protons, 
i.e. $\epsilon_e\la1$. We suggest alternative 
mechanisms that may be at work in relativistic collisionless shocks.

Magnetic fields are generated at shocks by the Weibel instability
\citep{ML99,CSA07}, that saturates its linear phase at a relatively 
low magnetic field, near the  equipartition with the electrons. 
At such low fields, protons keep streaming in current 
filaments, whereas the electrons, being much lighter than 
the protons, are quickly isotropized in the random fields 
and form a uniform background. Nonlinear evolution of the 
filaments leads to further amplification of the magnetic field up to 
$\epsilon_B\sim0.1$ on average (and $\epsilon_B\sim1$ locally in clumps)
as one approaches the main shock compression. 
The average Lorentz factor of the 
electrons is also gradually increasing toward the shock and
$\epsilon_e$ becomes 30\% to 50\% around and after the shock jump
(\citealp{CSA07}; Spitkovsky, in prep.). This electron
heating we are trying to understand.

\begin{figure}[t!]
\center\epsfig{width=4.1in,file=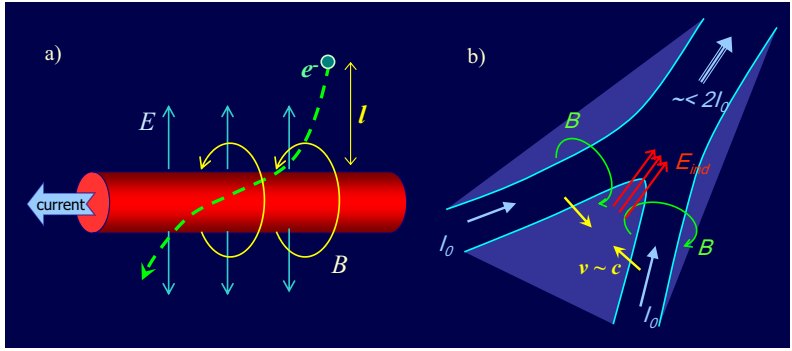}
\caption{ Acceleration mechanisms. (a) --- the electrostatic model, 
(b) --- the filament merger as an example of the induction model.  
\label{f0}} 
\end{figure}

{\it Electrostatic model} --- 
The current filaments are formed by the protons moving roughly at 
the speed of light (their Lorentz factor is $\sim\Gamma$). Hence, they  
are the sources of both the magnetic and electrostatic fields. 
In the absence of strong electrostatic shielding (simulations
seem to show this) $E$ and $B$ fields are related as $B\la E$. 
An electron, moving toward a filament, see Fig. \ref{f0}a, gains energy
$u_e\simeq e E l \simeq e B l$, where $l$ is the radial distance 
the electron travels through the strong $E$ field region. We normalize it to
the scaracteristic scale in the system, $c/\omega_{pp}$
(the typical size of the filaments is few $c/\omega_{pp}$)
as  $l\simeq\lambda(c/\omega_{pp})$,
where  $\omega_{pp}=(4\pi e^2 n/m_p \gamma_p)^{1/2}$
is the relativistic proton plasma frequency and $\gamma_p\simeq\Gamma$.
The parameter $\lambda$ accounts for the actual geometry of the filaments,
the electrostatic shielding in plasmas, the effects of the electrons
on the current distribution, etc. If most of the electrons reside
near the filaments (as simulations also seem to show), 
the electron energy density is estimated as
$U_e=n u_e\simeq n eB \lambda c/\omega_{pp}$, or in dimensionless units:
$\underline{\epsilon_e\simeq\lambda\sqrt{\epsilon_B}}$.
Although this mechanism does not provide the net acceleration (the
fields considered are potential), the electrons gain energy locally, 
and can radiate it away if the cooling timescale in filaments is short 
enough, see below.

{\it Induction model} --- 
The current filaments exhibit violent dynamics in which  
the field configuration can change rapidly; these include the filament 
multiple mergers (in the foreshock) 
and  break-up (mostly at the shock jump). Let's consider a merger 
as an example, Fig. \ref{f0}b. Two filaments with the typical magnetic 
field $B$ approaching each other with a velocity, $v\sim c$, 
induce the non-potential electric field $E_{\rm ind}$ in-between.
The typical size of the region with this field is of order
the filament size, hence it is $d\sim \delta (c/\omega_{pp})$, where $\delta$ 
is a dimensionless size of the merger region. An electron traversing the region
gains energy $u_e\sim e E_{\rm ind} d \sim e (v/c) B\delta c/\omega_{pp}$.
The corresponding electron energy density is $u_e n$. Thus, we obtain 
$\underline{\epsilon_e\simeq (v/c)\delta\sqrt{\epsilon_B}}$, which 
recovers the previous result once $\lambda$ is replaced with $(v/c)\delta$.
Perhaps not all plasma goes through such regions, so this mechnism
may explain acceleration of a smaller number of energetic electrons 
from the tail of the distribution.  
Unlike the electrostatic acceleration, the inductive acceleration
is ``permanent'', meaning that the electrons remain energetic 
in the downstream, where the current filaments are essentially gone.


{\bf Relevance to GRBs} ---
The radiative efficiency of a shock is determined by how fast the
bulk electrons lose their energy via radiation. If the electron
synchrotron cooling time is smaller than or comparable to the
electron residence time in the high field region (near the shock front), 
this electron will
radiate away energy comparable to its kinetic energy. The shock will
be radiatively efficient in this regime regardless of the field
dynamics in the far downstream. 
The dimensionless cooling time is defined as
$
T_{\rm cool}=t_{\rm cool} \omega_{pp}
=\left(\frac{6\pi m_e c}{\sigma_T\gamma_e B^2}\right)
\left(\frac{4\pi e^2 n'}{\Gamma m_p}\right)^{1/2},
$
where $t_{\rm cool}$ is the synchrotron cooling time,
$n'$ is the particle density behind the shock measured
in the downstream frame, $\Gamma$ is the shock Lorentz factor.
The region of strong magnetic field at the shock front, where 
$\epsilon_B\sim 0.1 - 0.05$ is of the size $d_B \sim50 c/\omega_{pp}$ or so.
Since the shock moves at $v=c/3$ in the downstream frame,
the residence time in the region of high field is
$t_{\rm res}\sim d_B/v\sim(150-300)\omega_{pp}^{-1}$.
This estimate, does not account for the filling factor of magnetic
inhomogeneities, which shortens the effective $t_{\rm res}$,
and the electron trapping in high-field clumps,
which is increasing $t_{\rm res}$.

If $t_{\rm cool}\la t_{\rm res}$, then the electrons lose
their energy quickly near the shock jump, hence the radiative
efficiency is high and such a shock can be seen as a GRB.
We refer to a shock as the {\it ``radiative shock''} if
$T_{\rm cool}\la300$ and as the {\it ``weakly radiative shock''} otherwise
(its efficiency depends on the fields in the far downstream, which 
have not yet been adequately probed in PIC simulations). In an extreme
case, $T_{\rm cool}<1$, called the {\it ``radiative foreshock''} regime, 
radiative cooling will be substantial even before the main shock compression,
hence cooling may change the entire shock structure.

Using the standard shock model \citep{RM05},
the comoving density behind an internal shock at a
radial distance $R$ from the central engine is
$
n=4\Gamma_i L/(4\pi R^2 \Gamma^2 m_p c^3)
$,
where $L$ is the kinetic luminosity, $\Gamma_i$ is $\Gamma$ 
of an internal shock. 
The magnetic field and the electron bulk Lorentz
factor are fractions $\epsilon_B$ and $\epsilon_e$
of the post-shock thermal energy density
$
B'=\left(8\pi\Gamma_i m_pc^2n'\epsilon_B\right)^{1/2}
$,
$
\gamma_e=(m_p/m_e)\Gamma_i\epsilon_e
$.
Parameters $\epsilon_B$ and $\epsilon_e$ are $\sim10\%$ and $\sim50\%$
as follows from simulations.
Finally, the dimensionless cooling time in
baryon-dominated internal shocks becomes
$\underline{
T_{\rm cool}^{(e^-p)}\simeq170 L_{52}^{-1/2}\Gamma_2\Gamma_i^{-3}
R_{12}\epsilon_B^{-1}\epsilon_e^{-1}}$,
Similarly, we evaluate the cooling time for the external (afterglow) shocks
for the constant density ISM and the Wind
($n\propto R^{-2}$) outflow models \citep{GPS99,CL00}; the details are in
(Medvedev \& Spitkovsky, in preparation).

\begin{figure}[t!]
\center
\epsfig{width=2.4in,file=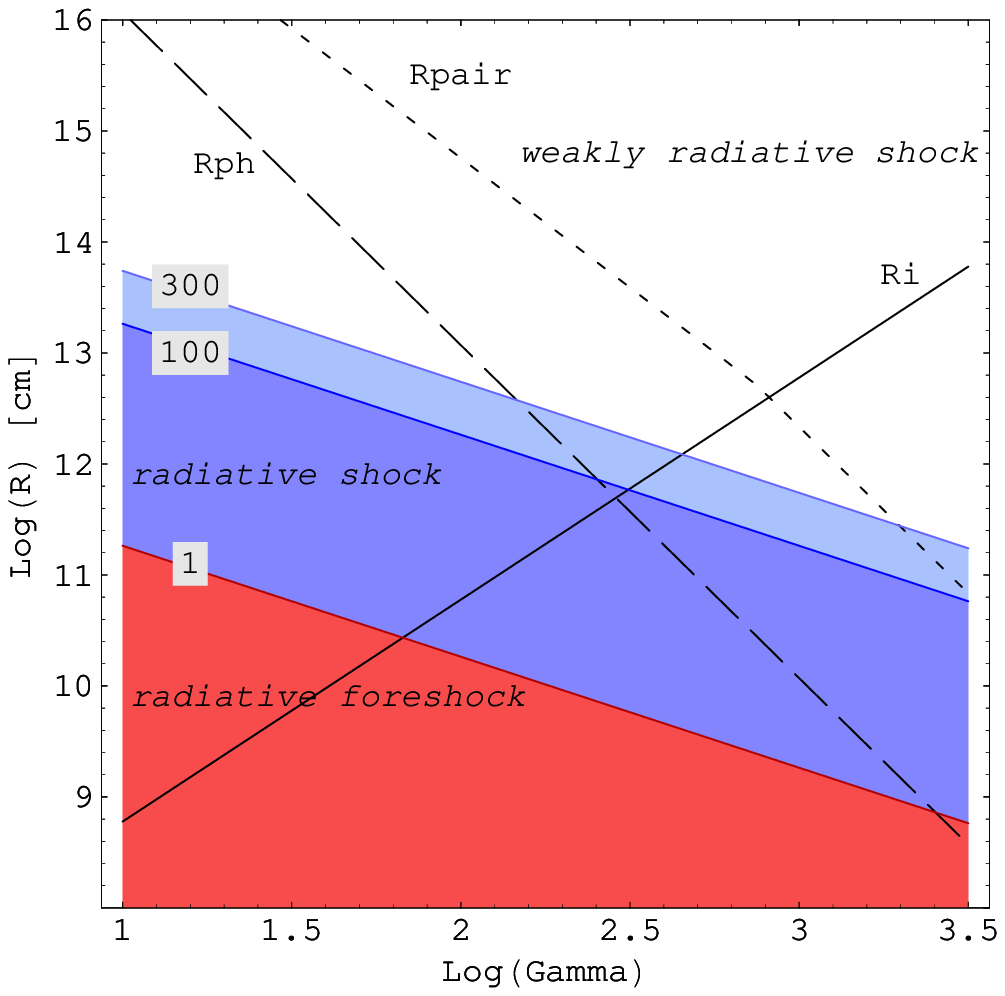}~~~%
\epsfig{width=2.4in,file=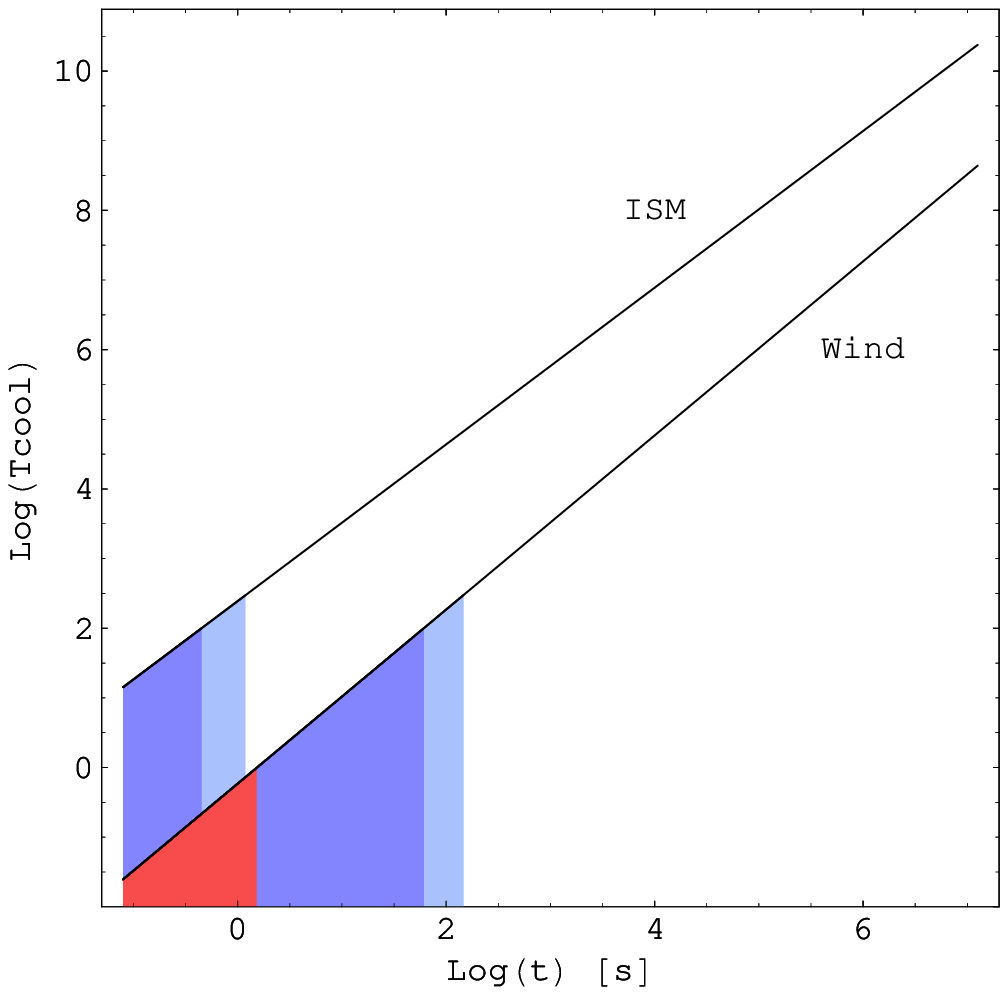}
\caption{ ({\it left}) --- Contours of $T_{\rm cool}$ vs. $\Gamma$
for the internal shocks for $T_{\rm cool}=1,\ 100,\ 300$.
Red filled regions correspond to
$T_{\rm cool}<1$ (radiative foreshock), dark and light blue
regions correspond to $1<T_{\rm cool}<100$ and $100<T_{\rm cool}<300$,
respectively (radiative shock), and the white
region corresponds to $T_{\rm cool}>300$ (weakly radiative shock).
Here $\Gamma_i=4$, $L_{52}=1,\ L_\gamma=0.1L,\ t_{v,-4}=1$.
We also mark the radii beyond which the internal shocks can form ($R_i$), 
the medium is optically thin to Thompson scattering ($R_{\rm ph}$)
and the optical depth due to $e^\pm$-pairs is below unity ($R_{\rm pair}$).
({\it right}) --- 
Cooling time in afterglows vs. time after the burst, for the ISM
and Wind models.
We use, $E=10^{54}$~erg, $A_*=10$ and $n_{\rm ISM}=100~{\rm cm}^{-3}$
and  a typical $z=2$.
\label{f1}}
\end{figure}

The results are shown in Fig. \ref{f1}.
One can see that internal shocks ($\Gamma_i\sim4$) with strong foreshock 
emission, $T_{\rm cool}\la1$, can occur for $\Gamma\la60$ and at
$R\la{\rm few}\times10^{10}-10^{11}$, well below the photosphere at
such low $\Gamma$'s. Strongly radiative shocks
$T_{\rm cool}\sim100-300$ can occur above the baryonic photosphere
in a relatively narrow, but very natural range of parameters, 
$\Gamma\sim150-400$, $\Gamma_i\ga2.5$, $R_i\sim10^{12\pm0.5}$cm and the outflow
kinetic luminosity $L\sim10^{52\pm2}$erg/s; hence they are likely observable.
Since $T_{\rm cool}\propto\Gamma_i^{-3}$, the region of the parameters 
widens greatly with increasing $\Gamma_i$.

For external shocks, we also see that, except for the very early times, 
the afterglow emission should be coming from far downstream,
not from the main shock compression region.
However, in the Wind model, the external shock can be radiative
up to $\sim100$~s after the burst, whereas for the ISM model, the radiative
shock regime can occur only at times earlier than a second after the explosion.
Since the afterglow usually sets in at least several
tens of seconds after the explosion (when enough external gas is
swept by the shock), we conclude that
very early afterglow emission can, in principle,
come from radiative shocks propagating in relatively strong Wolf-Rayet 
winds ($A_*\ga$~few).

It is remarkable that under these conditions, all the emission shall
come from a thin shell of thickness $\sim t_{\rm cool}c\sim1$ meter
(internal shocks) and $\sim100$ km (external shocks),
that is, from the region of main shock compression. This region
is already well resolved in 2D PIC simulations. Moreover, one
can obtain the emitted radiation directly from PIC simulations.

\end{document}